\documentclass{elsart}
\usepackage{graphicx}
\usepackage{hyperref}
\usepackage{natbib}
\usepackage{amsmath}

\journal{Ecological Complexity}
\def\erf{\mathop{\rm erf}\nolimits}  
\begin{document}

\title{Diffusion and Home Range Parameters for Rodents:
\emph{Peromyscus maniculatus}  in New Mexico}

\begin{frontmatter}

\author[consortium,cab]{G. Abramson},
\ead{abramson@cab.cnea.gov.ar}
\author[consortium]{L. Giuggioli},
\ead{giuggiol@unm.edu}
\author[consortium]{V. M. Kenkre}
\ead{kenkre@unm.edu}

\address[consortium]{Consortium of the Americas for Interdisciplinary Science,
University of New Mexico, Albuquerque, New Mexico 87131, USA.}

\address[cab]{Centro At\'{o}mico Bariloche, CONICET and Instituto Balseiro, 8400 San
Carlos de Bariloche, R\'{\i}o Negro, Argentina.}

\author{J. W. Dragoo},
\ead{jdragoo@unm.edu}
\author{R. R. Parmenter},
\ead{bparmenter@vallescaldera.gov}
\author{C. A. Parmenter},
\ead{cparment@sevilleta.unm.edu}
\author{T. L. Yates}
\ead{tyates@unm.edu}
\address{Department of Biology, University of New Mexico, Albuquerque,
New Mexico 87131, USA.}

\date{\today}

\begin{abstract}
We analyze data from a long term field project in New Mexico, consisting of
repeated sessions of mark-recaptures of \emph{Peromyscus maniculatus}
(Rodentia: Muridae), the host and reservoir of Sin Nombre Virus (Bunyaviridae:
Hantavirus). The displacements of the recaptured animals provide a means to
study their movement from a statistical point of view. We extract two
parameters from the data with the help of a simple model: the diffusion
constant of the rodents, and the size of their home range. The short time
behavior shows the motion to be approximately diffusive and the diffusion
constant to be $470 \pm 50$ m$^2$/day. The long time behavior provides an
estimation of the diameter of the rodent home ranges, with an average value of
$100 \pm 25$ m. As in previous investigations directed at \emph{Zygodontomys
brevicauda} observations in Panama, we use a box model for home range
estimation. We also use a harmonic model in the present investigation to study
the sensitivity of the conclusions to the model used and find that both models
lead to similar estimates.
\end{abstract}



\end{frontmatter}

\section{Introduction and Estimated Results}

Since the discovery of Sin Nombre Virus (Bunyaviridae: Hantavirus) in 1993 as
the agent of the severe Hantavirus Pulmonary Syndrome (HPS) in the North
American Southwest~\citep{nichol93}, a continuing effort is being maintained to
study the long-term dynamics of its principal host, the deer mouse,
\emph{Peromyscus maniculatus}~\citep{childs94}. As part of this effort, an
analytical model~\citep{abramson2002,abramson2003} has been developed and has
led to an understanding of spatio-temporal patterns that have been observed in
the field. A basic assumption of that model is that the rodent movement may be
regarded as diffusive and characterized by a parameter $D$, the diffusion
coefficient. Various contributions to the understanding of animal movement in
general~\citep{okubo,murray}, as well specific considerations for
Peromyscus~\citep{stickel,vessey}, form the background of this hypothesis. The
present paper is one in a series of investigations which have been launched to
assess the validity of the assumption of diffusive movement and to extract the
parameter $D$ from field observations.

In the first paper in the series~\citep{giuggioli2004}, hereafter referred to
as I, field data for \emph{Zygodontomys brevicauda} in Panama were used to
obtain $D$ from trapping measurements in a grid arrangement. During that
investigation, a procedure was given to deduce an \emph{additional} parameter
of the rodent motion: $L$ the home range size. It was found that $D\sim 200$
m$^2/$day and $L\sim 70$ m. Here we express the home range size as a length
rather than an area.

It is important to extract such parameters in different real scenarios. One
reason is to extend our knowledge about different animals involved in the
spread of the epidemic, the second to gain a thorough understanding of the
underlying theory. In the present paper, we investigate a system that differs
from that analyzed in I in three characteristics: the rodent, the region, and
the method of measurement. The rodent is \emph{Peromyscus maniculatus}, the
region is New Mexico (USA), and the measurement method is trapping in the web
arrangement \citep{parmenter03}. As will be seen below, we find for the present
system that short time measurements show the diffusion constant to be $D=470\pm
50$ m$^2/$day while the long time measurements show the home range size to be
$L=100\pm 25$ m\footnote{The error limits in the extracted diffusive
coefficient and the home range size, here and elsewhere in this paper, refer to
an estimate of all possible uncertainties. They include, among others, the
result of statistical error propagation and differences among the individuals
of the population.}. The latter is a consequence of using the box model
introduced in I for home range estimation. An additional question we raise in
the present paper concerns the sensitivity of the home range estimation to the
precise assumptions of the model. By using a harmonic potential (instead of a
box potential) to represent the return of the rodents to their burrows, we find
that the home range size is found  not to differ sizably from the box case.

\section{Trapping webs and the measurement of displacement}

The data we analyze in the present study correspond to field work performed in
the period from 1994 to 2003, at four sites in the state of New Mexico,
USA~\citep{mills99}. Eleven trapping webs were permanently set at each of the
four sites, and animals were captured on a monthly basis for three consecutive
nights on each occasion. Several species of \emph{Peromyscus} are present at
all sites, of which we chose \emph{P. maniculatus} for this analysis, both
because it was the most abundant and because it was the host and reservoir of
Sin Nombre virus. Restricting our attention to especially reliable data, we
found that the record contained 3765 captures of \emph{P. maniculatus},
corresponding to 1581 animals (794 males, 690 females, and 97 of unidentified
gender). The captures consisted mainly of adult animals, but juveniles and
sub-adults were captured as well. We performed the analysis only on adults
(1108 animals) which were also the ones that were recaptured most. A few young
animals were later recaptured as adults, but in general they were captured only
once; there is no way to know whether they were transient, or moved away from
the the current site to establish their home ranges elsewhere, or died sometime
after their initial capture event.

\begin{figure}[t]
\centering \resizebox{\columnwidth}{!}{\includegraphics{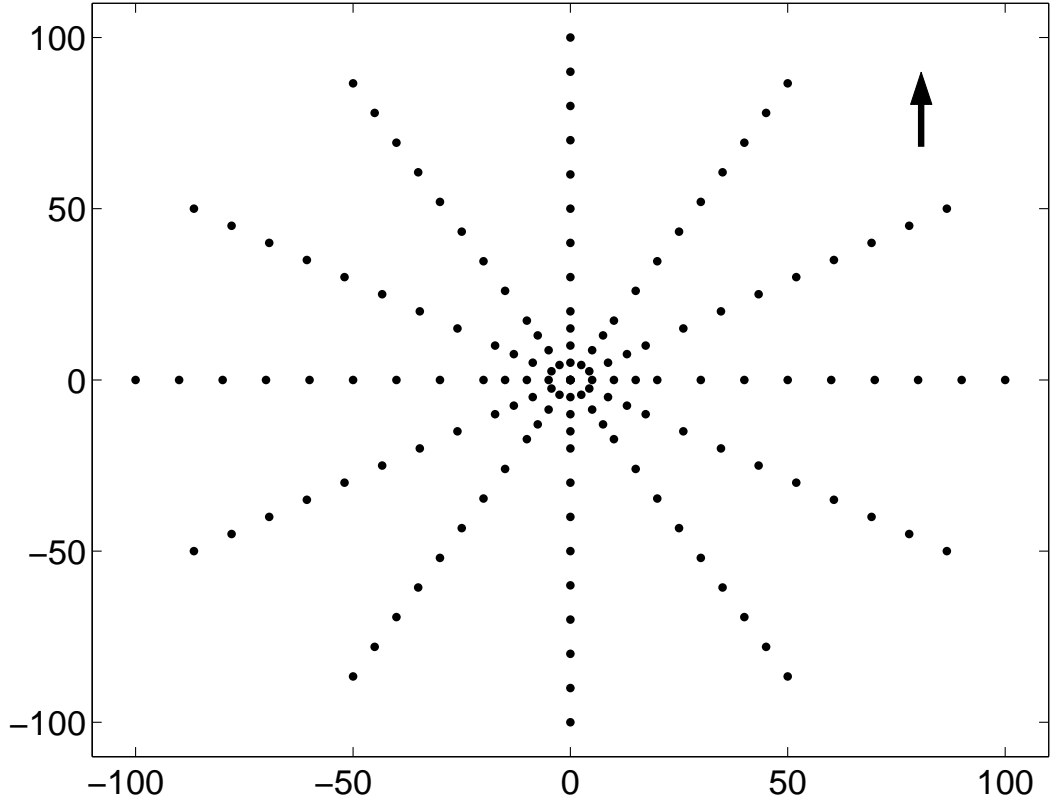}}
\caption{Arrangement of each of the 11 trapping webs (the arrow represents the
North direction). Each dot represents a
Sherman trap, except at the center, where there are four traps. The four inner
circles have radius increasing in 5 m intervals, while the rest are separated
by 10 m. (See \citep{parmenter03} for details.)}
\label{web}
\end{figure}

\begin{figure}[t]
\centering \resizebox{\columnwidth}{!}{\includegraphics{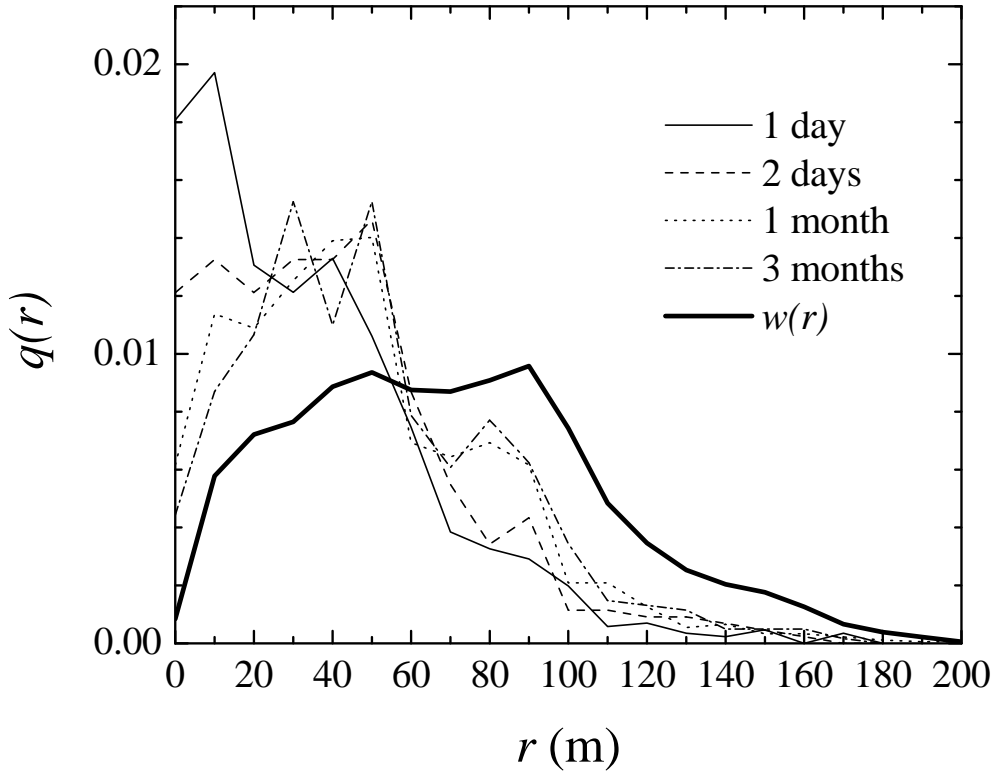}}
\caption{Observed distribution $q(r)$ of displacements $r$ at several time
scales, compared with the distribution $w(r)$ of distances in the web.}
\label{qsandw}
\end{figure}

The use of trapping webs such as the one schematized in Fig.~\ref{web}, as well
as the analysis of the capture-release-recapture observations, is an
implementation of the ``distance sampling'' method~\citep{buckland93}. It is
well documented and theoretically sound for the estimation of absolute
densities~\citep{parmenter03,anderson83}. For the study of the movement of the
animals, however, the web presents an obvious inconvenience: the distribution
of traps is very inhomogeneous. Certainly, it is precisely the inhomogeneity of
the design that is responsible for making this method particularly suited for
estimating the density at the center of the web. However, this design produces
a strong bias in the distribution of distances present in the array, as can be
observed in Fig.~\ref{qsandw} (heavy line, denoted $w(r)$).

Figure \ref{qsandw} also shows the observed distribution of the displacement of
mice at several time intervals, ranging from 1 day to 3 months, compared with
the distribution of distances in the web. Since the traps are measured monthly
during three consecutive days, we chose to sort the data according to the
following time intervals, that we refer to as ``time scales'': 1 day, 2 days,
and multiples of 30 days. As a result of the logistics of the field work, many
of the observed displacements do not correspond exactly to one of these time
intervals. We decided to assign those displacements to the closest of the time
scales. All the observations corresponding to each time scale are pooled
together, and the distributions shown in Fig.~\ref{qsandw} are their normalized
histograms.

It is apparent that, after just two days, the distribution $q(r)$ develops a
deep trough at $r=0$, reflecting the bias of the distances present in the web.
If the motion were purely diffusive, and if the measurements of the
displacements were fine enough, these distributions should be decreasing
Gaussians, always with a maximum at zero. The artifact produced by the web is
equivalent to a repulsion at short distances, as if the animals would prefer to
stay away from their initial position. This is obviously fictitious. It is
necessary to remove this artifact in order to obtain sensible measurements of
the motion. Nevertheless, the reader will note that, at 1 day, the distribution
of displacements still has a maximum at, or very near, zero. This is an
indication that the measurements on this time scale can be reliably used for
the estimation of the diffusion coefficient. Indeed, if one would draw
displacements at random, and uniformly, over the web, the observed distribution
would be that given by $w(r)$ in Fig.~\ref{qsandw}. Random displacements over
the web should be expected, for example, if at long times the diffusive
movement becomes restricted by the home ranges of the animals. That the
distribution at 1 day is far from this shape indicates that, on this time
scale, the saturation has not yet been attained.

\begin{figure}[t]
\centering \resizebox{\columnwidth}{!}{\includegraphics{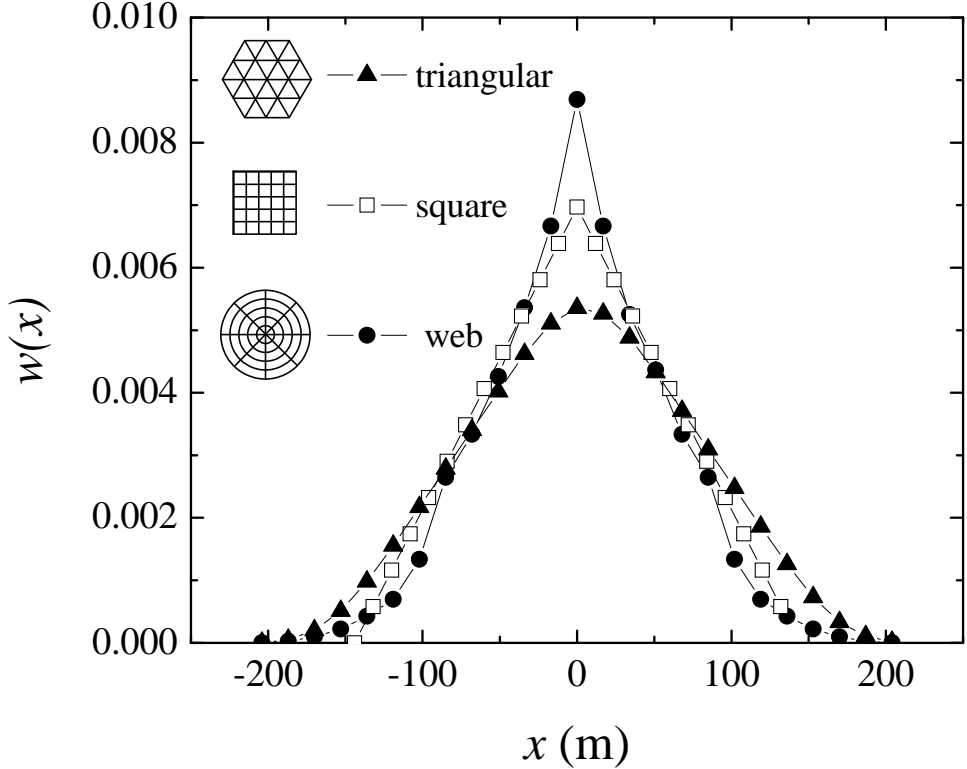}}
\caption{Distribution of East-West distances in three trapping configurations:
a web, a square grid, and a triangular grid (of hexagonal perimeter),
schematized in the legend. The diameter of the grids is 200 m in the three
cases, and the number of traps as approximately the same as it can be (given
the geometries of the designs).}
\label{grids}
\end{figure}

An alternative view is provided by Fig.~\ref{grids}. It shows the number of
distances present in the web in the East-West direction, as well as in two
alternative grids (discussed below). The over-sampling of short distances with
respect to long distances is very clear: an ideal device to measure
displacements should show a square distribution in this picture, homogeneous
for all distances up to the diameter of the region being sampled (in the
present case, a constant at 0.005, from --100 m to 100 m). It is worthwhile to
stress that the measurements have been taken with the purpose of monitoring the
density and its variations, and not the movement. This longitudinal study is,
nevertheless, so unique that it is worthwhile to use it to obtain other
demographic parameters such as, in this case, the diffusion ones.

Let us consider alternatives to the web of traps, and how they might affect the
measurement of the displacements. Figure~\ref{grids} shows a comparison of the
distribution of distances in the East-West direction for three trapping
configurations: a web (as in Fig.\ref{web}), a square lattice, and a triangular
lattice of hexagonal perimeter (see the simplified diagrams in the legend of
Fig.~\ref{grids}.) For the purpose of a sensible comparison, the three
configurations have all been taken to have a diameter of 200 m, understood as
the longest distance within each configuration, but the number of traps has
been taken different in each case. The number selected in each case
approximates, as best as possible given the geometrical constrains, the 148
traps used in the webs of Fig.~\ref{web}: 144 for the square grid, and 127 for
the triangular grid. It is easy to see that, of the three, the web has the
least homogeneous distribution of distances, with a sharp peak at zero. To
quantify this effect, we calculate the fraction $f$ of distances smaller than
one fourth the diameter, viz. 50 m in this case. The results for the three
configurations are as follow:
\[
f_{\mbox{\small web}} = 0.63,~~f_{\mbox{\small sq.}} = 0.56,~~f_{\mbox{\small
tr.}} = 0.50.
\]
The triangular grid is not only the one with the smallest $f$ (thus closer to
the value a uniform distribution would have) but it also has a flat maximum at
zero. A sufficiently large triangular lattice could provide a wide range of
distances with a rather flat distribution, and thus could be a better choice
for the measurement of displacements. A larger trapping web would comprise,
however, more traps and the possibility of its use should be assessed for each
intended application.

\section{Renormalization of the measurements and estimation of the diffusion constants}

\label{diffusion}

We use the discussion of the previous section in the analysis of the
displacements. Let us consider the displacements of recaptured animals as a
statistical ensemble, representing the movement of a hypothetical
mouse---henceforth referred to as a ``walker''---whose statistical properties
we intend to derive from the data. When observed on a short time scale, the
motion of the walker might be approximately diffusive. At longer times, both
the existence of home ranges and the finiteness of the array should take over,
constraining the walk. Other components of the mouse movement, such as
explorations and shifting of the home range, might appear as well, but since
they are more rare and we are averaging over an ensemble of animals, we
restrict ourselves to the consideration of confined diffusive motion.

\begin{figure}[t]
\centering \resizebox{\columnwidth}{!}{\includegraphics{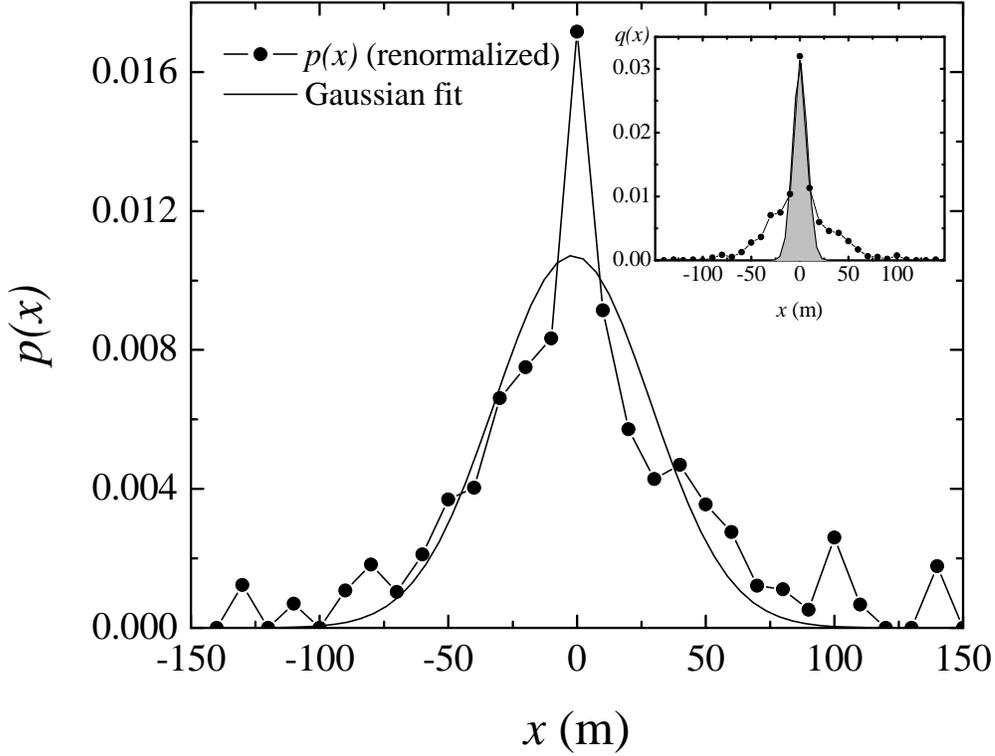}}
\caption{Distribution of displacements in the East-West direction at 1 day,
renormalized with the distribution of distances $w(x)$. The continuous line
shows the least squares Gaussian fit, and captures well ($\chi^2 =1.3\times
10^{-4}$) the width of the distribution. The inset shows the distribution
obtained directly from the data, before the renormalization, together with the
least squares Gaussian fit (in grey, $\chi^2 =2.2\times 10^{-4}$). The
``shoulder'' arising from the existence of the inner and finer set of traps,
and the failure of the Gaussian to represent this shoulder, are both clearly
apparent.}
\label{q1day}
\end{figure}

Supposing that the motion of the mice is indeed diffusive, the probability
density function of the displacements of an ensemble of mice from an initial
position is just the standard (Gaussian) propagator of diffusive motion, with
diffusion coefficients that are generally different  in different directions
because of anisotropy in the motion. If the motion of the mice could be
followed with arbitrary precision, one would expect the mean squared
displacement to be linear in time at short times, before saturation takes over
to limit the size of the displacements. However, the measurements of position
are taken with a discrete device. The corresponding values of the probability
density are, therefore, biased. It is possible to take into account the
distribution of distances between traps in the web to
have the effect of this bias, in the following way. The probability $%
Q(x)=q(x)dx$ of observing a displacement between $x$ and $x+dx$, is equal to
the probability $P(x)=p(x)dx$ that, in a day's time, the walker actually makes
such a movement, multiplied by the probability that the web contains such a
distance, $W(x)=w(x)dx$. We can renormalize the observations to obtain the
distribution of displacements that characterized the movement:
\begin{equation}
P(x) = \frac{Q(x)}{W(x)}=\frac{q(x)}{w(x)}.
\end{equation}

The distributions $q(x)$ and $w(x)$ can be built from the recapture data and
from the geometry of the web, respectively ($w(x)$ has indeed been shown in
Fig.~\ref{grids}). As expected, the distribution $p(x)$ is bell-shaped, as can
be observed in Fig.~\ref{q1day}. Moreover, on the 1-day time scale, it is well
fitted by a Gaussian ($\chi^2 =1.3\times 10^{-4}$), supporting the hypothesis
that the movement is initially diffusive. Remarkably, the distribution $q(x)$,
shown as an inset in Fig.~\ref{q1day}, while bell-shaped, cannot be well
represented by a Gaussian.

Within the approximation that the movement is initially diffusive, we can
identify this Gaussian with the propagator of the diffusion process at 1 day,
which is the shortest time scale available from the measurements. The result,
in both the $x$ (East-West) and the $y$ (North-South) directions, is:
\begin{eqnarray}
D_x &=& 460 \pm 50 \mbox{ m$^2$/day}, \\
D_y &=& 490 \pm 50 \mbox{ m$^2$/day},
\end{eqnarray}
or an average of $470 \pm 50$ m$^2$/day. A typical distance covered diffusively
in one day is, thus, of the order of 20 m.

\section{Saturation of the mean square displacement and evaluation of home ranges}

\begin{figure}[t]
\centering  \resizebox{\columnwidth}{!}{\includegraphics{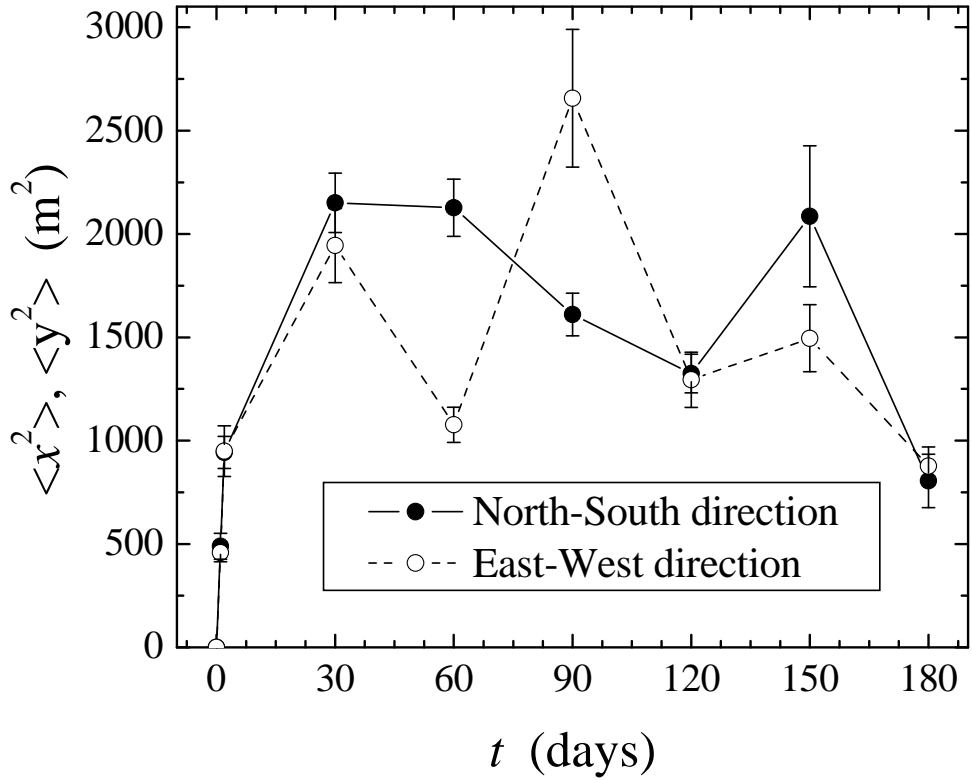}}
\caption{Mean square displacement as a function of time, in the two directions.
The values are calculated as the second moment of the  Gaussians fitted to the
renormalized distributions of displacements. The initial  slope of each curve
is $2Dt$.}
\label{xsquare}
\end{figure}

The analysis described in Section~\ref{diffusion} can be carried over at each
of the accessible time scales: over a day, two days, and from one  month to six
months. The result is shown in Fig.~\ref{xsquare}, where the  growth of the
mean square displacement of the walker (in both directions) is  shown as a
function of time. After the initial linear regime the curves  saturate (with
strong fluctuations due to indeterminancies of the measurements)  in both
directions to the values
\begin{eqnarray}
\left\langle x^{2}\right\rangle &=&1560\pm 650\mbox{ m}^{2},
\nonumber \\
\left\langle y^{2}\right\rangle &=&1680\pm 540\mbox{ m}^{2}.
\label{saturation}
\end{eqnarray}

As explained in I, this saturation is the result of a combination of two
effects: the existence of home ranges on the one hand and of a spatial window
of available observations---provided by the trapping array---on  the other. It
is easy to envisage two simple models of rodent motion. In  one, the existence
of the home range is represented by motion that is  confined within a finite
area but assumed to be otherwise free (and of course diffusive as is
appropriate to a random walker). We call this the box model. In the other model
the attraction the rodent feels to its burrow  is represented by a potential
typically continuous in space and the motion  is described~\citep{kk} by a Fokker-Planck
equation~\citep{risken} instead of the simple diffusion equation. The simple
assumption that the potential is harmonic yields what we call the harmonic
model.

\begin{figure}[t]
\centering \resizebox{\columnwidth}{!}{\includegraphics{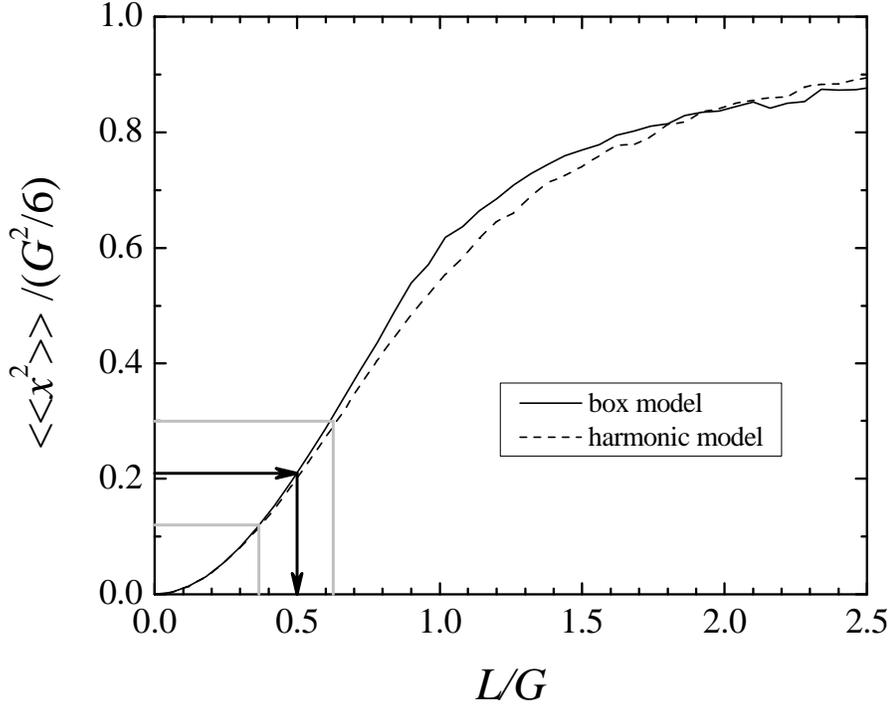}}
\caption{Simulated mean square displacement for the two potentials, box (solid
line) and harmonic (dashed line). The arrows show the value corresponding to
the field measurements of {\em P. maniculatus}, and the consequent value of the
home range size, as $L=100\pm 25$ m. The grey lines on the sides of the arrows
delimit the error interval. The notation $\langle\langle x^2\rangle\rangle$
indicates that an additional average (over the burrow positions $x_c$) has been
performed on $\langle x^2_{sat}\rangle$ of Eq.~(\ref{gmsd}).}
\label{xsquare-model}
\end{figure}

While we mentioned both these models in our earlier work~\citep{giuggioli2004},
we used only one of them, the box model, for parameter estimation.  The box
model procedure to obtain the home range size consists of evaluating the mean
square displacement at steady state inside the grid of width $G$ considering
that each captured-recaptured mouse has a home range that  could be centered
anywhere in space. We perform numerical simulations to  represent such a real
capture-recapture experiment. A typical simulation consists  of randomly taking
two positions $x_{0}$ and $x_{1}$ from a uniform distribution of width $L$
centered at $x_{c}$ in space. Only those  values such that the displacement $x$
is completely contained in the  window $(-G/2,G/2)$ are considered actual
observations, and stored for the calculation of the mean square displacement
$\langle x^{2}\rangle $. The mean square displacements are then averaged over
the burrow  position $x_c$, characterizing each mouse. The resultant plot as a
function of $L$ (relative to $G$) is displayed in Fig. \ref{xsquare-model}.

With this procedure it is possible to calculate, with the aid of the simulated
result, the value of the home range for \emph{P. maniculatus} in New Mexico.
This is shown by the arrows in Fig.~\ref{xsquare-model} to be $L=100\pm 25$ m.
This value is a little smaller  than those that correspond to the report of
Stickel [see~\citep{stickel}, p. 380]. The latter are expressed in units of
area: that averages $1.77$ ha (with a range from $0.67$ to $4.02$ ha), for a
single observation of \emph{P. m. blandus} during the summer, in a mesquite
habitat in New Mexico.

We now ask what the sensitivity of these results is to the assumed square shape
of the potential in the box model. If the potential is, instead, harmonic, the
simulation procedure changes only with respect to the distribution of initial
capture and recapture: it is now Gaussian rather  than uniform. To allow a
meaningful comparison, we choose the width of the Gaussian distribution as
being related to the confining length in the box model in such a way that
results of both agree in the limit $\zeta =G/L\rightarrow
+\infty $. Here $\zeta $ is the ratio of the grid size $G$ to the
characteristic confining length $L$. In the box case $L$ equals the width of
the box. In the harmonic case it is simply related to the curvature of the
harmonic potential. In the opposite limit, $\zeta \rightarrow 0$, the mean
square displacement is the same in both models. The determination of $L$ for
the harmonic model proceeds as follows.

The mean square displacement at long times for a mouse roaming inside  its home
range, concentric with the grid, is given by the quantity
\begin{equation}
\left\langle x_{sat}^{2}\right\rangle  =\frac{\int_{-G/2}^{G/2}dx_{0}%
\int_{-G/2}^{G/2}dx_1\left( x_1-x_{0}\right)  ^{2}\mathcal{P}_{st}(x_{0})
\mathcal{P}_{st}(x_1)}{\int_{-G/2}^{G/2}dx_{0}\int_{-G/2}^{G/ 2}dx_1
\mathcal{P}_{st}(x_{0})\mathcal{P}_{st}(x_1)},  \label{gmsd}
\end{equation}
where $\mathcal{P}_{st}(x)=\exp \left[ -U(x) /D\right]  \left\{
\int_{-\infty }^{+\infty }dx\exp \left[ -U(x) /D\right]
\right\}^{-1}$ is the stationary distribution of the Fokker-Planck equation for
the potential $U(x)$. In the harmonic case $U(x)=\gamma x^{2}/2$, and the
evaluation of Eq. (\ref{gmsd}) gives

\begin{equation}
\left\langle x_{sat}^{2}\right\rangle =\frac{2D}{\gamma }\left[  1-%
\frac{2Ge^{-\frac{G^{2}}{8D/\gamma}}}{%
\sqrt{8\pi D/\gamma }\erf\left( G/\sqrt{8D/\gamma }\right) }\right] ,
\end{equation}
which equals $G^{2}/6$ when $G/(D/\gamma )\ll 1$ and $2D/\gamma$ when
$G/(D/\gamma )\gg 1$. Requiring that the limit of $\left\langle
x_{sat}^{2}\right\rangle $  when  $G\rightarrow +\infty $ is equal for both
potentials, implies that $L=\sqrt{12D/\gamma }$. As a consequence, the
stationary distribution for the harmonic model is
\begin{equation}
\mathcal{P}_{st}(x)=\frac{e^{-\frac{x2}{L^2/6}}}{L\sqrt{\pi/6}}.
\label{pst-harm}
\end{equation}

Using expression (\ref{pst-harm}) we have performed a numerical simulation of
the harmonic model, which is shown in Fig.~\ref{xsquare-model} as a dashed
line. For the measured mean square displacement (marked with arrows in the
plot) the result is practically indistinguishable from that of the box model,
as is visually clear from the near-coincidence of the solid and the dashed
lines in Fig.~\ref{xsquare-model}.

For this set of data we thus see that the results are not particularly
dependent on the potential used. It is however important to determine  the
sensitivity of the predicted mean square displacement to the choice of  the
model potential in a general way, considering arbitrary potentials, and  also
carrying out the analysis for temporal evolution as well as for  saturation
values. We have undertaken such a detailed analysis and its results  will be
reported elsewhere~\citep{giuggioli2004b}.

\section{Discussion}

Underlying the spatio-temporal behavior of the hantavirus epizootic there are
numerous mechanisms, some ecological, some epidemiological, some environmental,
and still some of non-biological nature, such as meteorological disruptions
like El Ni\~{n}o. The dynamical model~\citep{abramson2002} that forms the basis of
our recent theoretical investigations~\citep{abramson2003} requires, for its
quantitative application, knowledge of rodent parameters, in particular the
diffusion constant. We have set out in I a procedure to evaluate the diffusion
constant from recorded rodent movement in grid-based traps and continued that
evaluation in the present paper for web-based traps. This latter set of data
has been collected from the long term longitudinal study carried out at the
University of New Mexico since shortly after the outbreak of HPS, at various
places in New Mexico. We have used the displacements of the mice within the
capture webs to derive the statistical properties of their movement. This
analysis has provided us not only with an estimation of the diffusion
coefficient, but also with information about the home range of the mice.

For our present calculations, the displacement data had to be renormalized with
respect to the distribution of distances of the capture web. Indeed, since the
webs have been designed to measure absolute densities of the population, and
not for the measurement of individual displacement, this distribution is very
inhomogeneous. We have shown, however, that it is possible to compensate for
this inhomogeneity to obtain a good estimate of the actual distribution of
displacements of the population. We have shown, besides, how different
geometrical designs of the trapping grid would affect the measurements in
different ways.

Extraction of the diffusion constant $D$ from the short time observations is
straightforward, has been explained in I, and yields $D$ for \emph{P.
maniculatus} in New Mexico to be 470 m$^2$/day, which is larger by a factor of
at least 2 relative to the $D$ deduced in I for \emph{Z. brevicauda} in Panama.
This is perhaps in keeping with the expectation that the desert environment
requires the mice to move faster and farther to procure the more scant
resources.

The observed saturation of the mean square displacement implies the existence
of a length scale. There are two separate length scales in the system analyzed.
One of them is imposed by the measurement and is the size of the trap region,
denoted by $G$ in our analysis. The other is a true characteristic of the
rodent motion, the home range size $L$. It is important to distinguish between
the effect of these two separate quantities on the observations. Our
theoretical studies have shown that the saturation value of the mean square
displacement varies in the form of a sigmoidal curve as a function of the ratio
$L/G$. For grid sizes large with respect to the home range, the mean square
displacement varies as the square of the home range while in the opposite limit
of small grid size it is proportional to the square of the grid size. Not only
is this behavior qualitatively similar for the box model and the harmonic model
but the deduced value of the home range is also quantitatively similar as we
have shown in Fig.~\ref{xsquare-model}. The question of the precise influence
that the assumed model potential has on the interpretation poses an interesting
theoretical challenge. Our ongoing work on that issue addresses arbitrary
potentials and will be reported elsewhere~\citep{giuggioli2004b}.

While the initial purpose of our investigations was the estimation of the
diffusion constant of the mice from field observations, the analysis in I as
well as in the present paper has also resulted in attention being focused on
the concept of home ranges. They have, of course, been known and discussed
widely earlier in the literature
\citep{burt43,stickel,anderson82,ford74,wolff89,wolff03}. Our considerations
make clear that it is crucial to incorporate them in descriptions of the spread
of epidemics such as the Hantavirus. We are currently working towards the
development of a series of new theoretical models for this purpose. One of the
underlying ideas in the new models is to consider some of the mice to be
largely stationary (moving within their home ranges) and others to be largely
itinerant (in search of home ranges of their own)~\citep{kenkre04}.  Another
important alternative is to consider the effects of gender on the motion and
home ranges~\citep{wolff92,wolff97} and of the dependence of the home range
size on the density of the rodents~\citep{parmenter83,wolff84,wolff03}. We are
trying to assess information about such issues from data available to us from
the New Mexico observations.

\begin{ack}
It is a pleasure to thank Aaron Denney for discussions. This work was supported
in part by the NSF under grant no. INT-0336343, by NSF/NIH Ecology of
Infectious Diseases under grant no. EF-0326757, by the CDC Cooperative
Agreement no. USO/CCU613416-01, and by DARPA under grant no.
DARPA-N00014-03-1-0900. G. Abramson acknowledges partial funding by CONICET
(PEI 6482), and by Fundaci\'{o}n Antorchas.
\end{ack}

\end{document}